# Heat Generation in Spatially Confined Solids Through Electronic Light Scattering

S.S. KHARINTSEV[1*] AND E.I. BATTALOVA[1]

[1]*Department of Optics and Nanophotonics, Institute of Physics, Kazan Federal University, Kremlevskaya str., 16a, Kazan, 420008, Russia*
*[*skharint@gmail.com](mailto:skharint@gmail.com)*

**Abstract:** In this *Letter*, we study the optical heating of spatially dispersive solids due to electronic light scattering (ELS), a phenomenon driven by indirect optical transitions. In this process, a spatial heterogeneity generates an optical near-field photon with expanded momentum, leading to electron-photon momentum matching, followed by thermalization of the electron system promoting optical heating. We experimentally demonstrate this effect by melting spatially confined semiconductor (Si) and metal (Au) under continuous-wave laser pump within the transparency window. ELS is a dominant physical process in media with strong spatial dispersion, underlying a variety of thermo-optical applications.

Everyday intuition suggests that optical heating of materials is directly associated with light absorption. This physical process plays a paramount role in most areas of photonics and optical spectroscopy.[1–5] Though this phenomenon is well-studied for homogeneous materials, the spatial non-locality relevant to heterogeneous media remains overlooked by the scientific society so far.[6] The real refractive index $n$ and the extinction coefficient $\kappa$ are fundamental optical constants driving scattering and absorption of light, respectively (Fig. 1a).[7–9] Both mechanisms can perturb an equilibrium of the electron system, leading to spatial-temporal fluctuations of charge density $\rho(\boldsymbol{r},t)=\rho_0+\frac{1}{e}\nabla\cdot\boldsymbol{P}(\boldsymbol{r},t)$ ($\rho_0$ is the equilibrium charge density, $e$ is the electron charge, $\boldsymbol{P}(\boldsymbol{r},t)$ is the electric polarization of medium).[10] Following M.V. Klein,[11] the cycle-averaged charge density $\langle\rho(\boldsymbol{r},t)\rangle$ corresponds to a true phase-delay refractive index of homogeneous media in the lack of perturbation, i.e. $n\sim\rho_0$. Since light absorption is a process that is nonlocal in time, the change in $\langle\rho(\boldsymbol{r},t)\rangle$ and $n$ is associated with temporal dispersion at resonance. The causality principle not only relates $n$ and $\kappa$ to each other via the Kramers-Kronig formula but also imposes the limitations on them.[12] It has long been believed that the refractive index of naturally occurring materials cannot exceed 4.[7,9] Recent experimental works have shown that these limitations are no longer valid. The underlying physical mechanisms include gap plasmon resonance excitation in self-assembled gold nanoparticles (Au NPs) ($n = 10.02$),[4] phase transitions in perovskites ($n = 26$),[13] infrared plasmon-polariton excitation in a graphene monolayer near a quantum well ($n > 100$).[14] Remarkably, I. Kaminer et al introduced a momentum-dependent confinement factor instead of the phase-delay refractive index that is commonly accepted.[14,15] The authors highlighted the emerging primacy of spatial non-locality in media involving an optical inhomogeneity whose size is comparable to or even less than the wavelength of an incident electromagnetic wave. The generation of the near-field in close vicinity of the optical inhomogeneity leads to spatially varying refractive index $n(\boldsymbol{r})$ or spatial dispersion $\nabla[n(\boldsymbol{r})] = dn/d\boldsymbol{r}$.[6,16] The interaction of light with spatially dispersive matter forms an emerging field of modern optics, called *nonlocal photonics*.[16]

This *Letter* focuses on the optical heating of spatially dispersive solids due to electronic light scattering (ELS), a phenomenon close to electronic Raman scattering in which initial and final electronic states are different, and optical transitions can be indirect due to electron-photon momentum matching. This is specifically the physical mechanism that unravels a number of unusual optical and spectroscopic phenomena that are still poorly understood. These include the optical melting of spatially confined solids under continuous-wave sub-band pumping[17] and the nonlinear increase in blue-shifted Raman intensity upon optical heating.[18,19]

Strong spatial dispersion occurs, for example, in a crystal-liquid dual system exhibiting long-range order and local disorder that coexist.[6,20] This class includes ceramics and high-entropy crystals,[21] amorphous and porous solids,[22] perovskites,[23] liquid crystals,[21] highly-associated liquids,[24] intercellular water,[25] hydrogels,[26] peptides and proteins,[27] to name a few.

For optically transparent ($\kappa \approx 0$) semiconductors the spatially varying refractive index reads as

$$n^2(\boldsymbol{r}) = 1 + \frac{e^2}{\pi^2 m_e} \sum_{cv} \int_{BZ} \frac{f_{cv}^{\varkappa}(\boldsymbol{k})}{\Omega_{cv}^2(\boldsymbol{k}) - \omega^2} d\boldsymbol{k}, \quad (1)$$

where $\Omega_{cv}$ is a vibronic frequency corresponding to optical transitions between the Bloch electronic states $|v\rangle$ and $|c\rangle$, $m_e$ is the effective mass of an electron, $\boldsymbol{\varkappa}$ is a light polarization direction. The oscillator strength $f_{cv}^{\varkappa}(\boldsymbol{k})$ is modified as follows

$$f_{cv}^{\varkappa}(\boldsymbol{k}) = \frac{2m_e}{\hbar} |D_{cv}^{\varkappa}(\boldsymbol{k})|^2 \quad (2)$$

here $\hbar$ is the Planck's constant, $D_{cv}^{\varkappa}(\boldsymbol{k}) = \langle c | e^{k(r)\varkappa r} \, \partial/\partial \boldsymbol{r} \, | v \rangle$ is the transient electrical dipole moment that considers spatial dispersion. In homogeneous media, a change in the refractive index is possible at resonance only (temporal dispersion) through vertical (or direct) transitions, leading to absorption (Fig. 1a). However, the refractive index is resistant to vibrational Raman scattering (VRS) because the electron system remains unperturbed (Fig. 1b). In Eq. (1), integration runs over the entire Brillouin zone, and, thus, indirect optical transitions become

accessible and the real refractive index can increase significantly. It results in ELS, resembling electronic Raman effect in which the electron momentum can change (Fig. 1c).[5] ELS is a specific signature of nonlocal media. Due to this physical mechanism the real refractive index increases with increasing charge carriers concentration in the conduction band.[3] Depending on the energy band landscape, this process can lead to heating of a light-scattering medium through electron-phonon interaction (Fig. 1c).

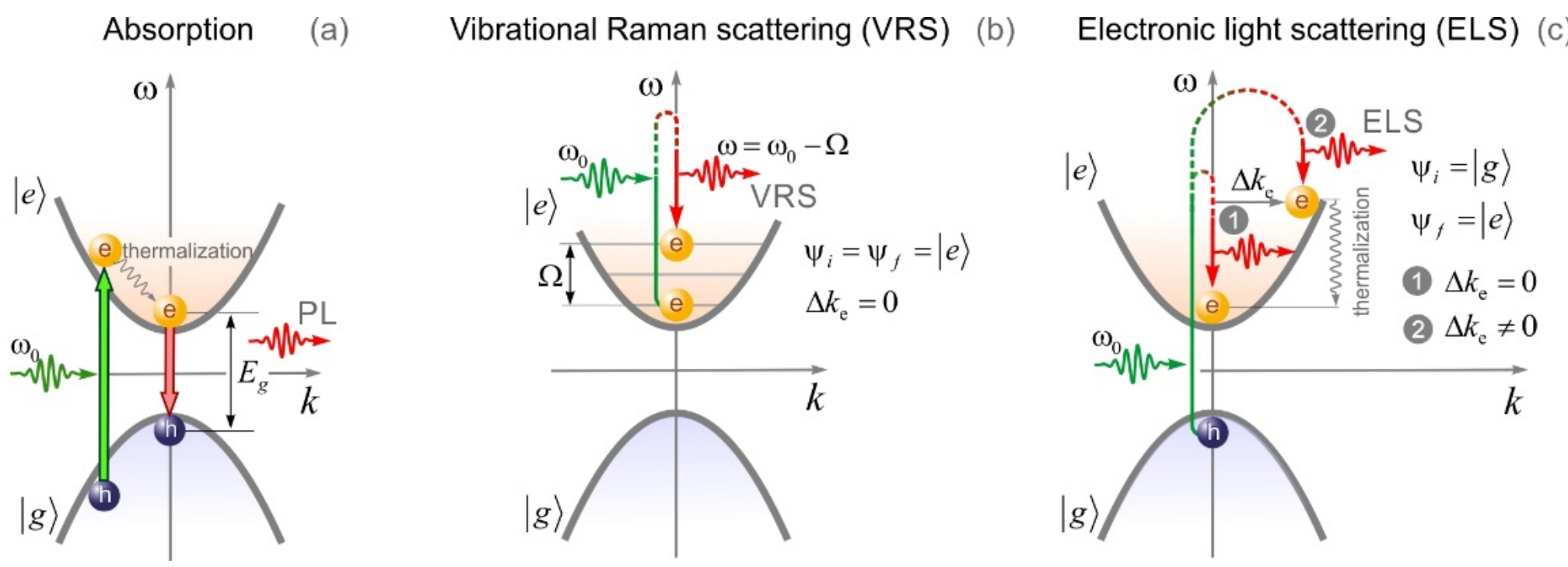

**Figure 1.** Light-matter interaction mechanisms in semiconductors: (a) absorption, (b) vibrational Raman scattering (Raman effect) and (c) electronic light scattering without and with a change in the electron momentum.

Optical heating of a temporal-spatial causal medium with a complex permittivity $\varepsilon(\omega,k) = \varepsilon'(\omega,k) + i\varepsilon''(\omega,k)$ ($\varepsilon'$ and $\varepsilon''$ are the real and imaginary parts of permittivity) is governed by the absorbed power in volume $V$:

$$P_{abs} = \int_V \omega\varepsilon_0\varepsilon''(\omega,k)|\boldsymbol{E}(\omega,k)|^2 dV, \tag{3}$$

where $|\boldsymbol{E}(\omega,k)|^2$ is the electric field intensity. Since $\varepsilon'' = 2n\kappa$,[28] there are two contributions to material optical loss: 1) resonant absorption ($\kappa$) and 2) inelastic free-electron Drude scattering ($n$).[29] Light is absorbed by the medium through direct optical transitions between the electronic states near the bandgap. Another mechanism is associated with the attenuation of light intensity due to its inelastic scattering by free charges. The attenuation of the intensity of the incoming light $I_0$ when it passes through a homogeneous medium over a distance $x$ exceeding the wavelength $\lambda$, is governed by the Beer-Lambert law $I(x) = I_0\exp[-\alpha x]$, wherein the absorption coefficient $\alpha = 2\omega\kappa/c$ ($c$ is the speed of light in vacuum)[28] is solely contributed by $\kappa$. Clearly, this formalism is insufficient for understanding an anomalous increase in light absorption (two orders of magnitude) by a Si layer only 2 nm thick, previously discussed in Ref.[17] In this work, the authors made conclusions based on the optical melting of a Si AFM cantilever apex (the melting point is 1687 K for bulk Si) and depleted reflection of Si bulk decorated with 1-3 nm diameter Au NPs over all energies above the bandgap of 1.1 eV. However, there are still no solid arguments in favor of absorption driven by indirect transitions under non-resonant pumping. Though this process is also possible,[30,31] its efficiency is unlikely to be high because of energy band $k$-dispersion.

The $k$-dependent ELS intensity is determined by the photonic density of states $\rho(k)$ and the population of electrons, driven by Fermi-Dirac statistics $f_{FD}(k)$, namely:

$$I_{ELS}(k) = C \int \rho(k-k')E(k')f_{FD}(k')dk', \tag{4}$$

where $C$ is a constant proportional the scattering cross-section, $E(k) = \hbar(\omega_0 - \Omega(k))$ is the scattered photon energy. In the simplest case when $\rho(k-k') = I_0\delta(k-k')$ ($I_0$ is the incident intensity, $\delta(k-k')$ is the Dirac-delta function), one gets

$$I_{ELS}(k) = C\frac{E(k)}{\mathrm{e}^{\frac{E(k)-E_p}{k_BT}}+1}I_0 \cong CE(k)\mathrm{e}^{-\frac{E(k)-E_p}{k_BT}}I_0, \tag{5}$$

where $k_B$ is the Boltzmann constant, $T$ is a temperature. This formula serves as a good approximation provided that high spatial homogeneity $\rho(k)\sim\delta(k)$ and $E(k) - E_p \gg k_BT$. In Eq. (5), we introduced an additional parameter $E_p$ analogous to Penn energy.[32] This seminal concept specifies the energy at which the electronic density of states is maximum throughout the entire Brillouin zone.[9] In our model, this parameter $E_p = \hbar\omega_0 - \hbar\Omega(k_p)$ denotes the frequency shift corresponding to the statistical mode of near-field photon momentum $k_p$. This shift tends to zero for electrons trapped near the conduction band edge. Such transitions basically contribute to the appearance of the low-energy ELS peak. This is the reason why the central peak is dominant in spatially confined metals where indirect transitions near the Fermi level occur.

Evidently, high-frequency ELS leads to optical heating due to $k$-dispersion and electron thermalization. This means that the optical heating is determined by the slope of the energy band $dE/dk$ and the population of electrons governed by Fermi-Dirac statistics. Thus, the change in overall temperature without heat transfer to the environment can be estimated using the following formula:

$$\delta T = \frac{1}{k_B}\int_{BZ}\frac{dE(k\prime)}{dk\prime}f_{FD}(k')dk'. \tag{6}$$

Eq. (6) indicates the fact that solids with the flat energy band valley are not heated under sub-band pump, and vice versa.

Fig. 2a schematically shows a TERS setup for measuring the optical heating of spatially confined Si, earlier suggested in Ref.[17] In our experiment, a Si AFM cantilever (VIT_P, NT-MDT) top-illuminated by a focused 633 nm laser light (x100, N.A.=0.7) oscillates in semicontact mode over a borosilicate glass substrate covered with a 50 nm Au film (TED PELLA, Inc.). We first scan the film surface with the AFM cantilever to visualize its roughness. The size of surface irregularities ranges from 1 to 2 nm (Section I, the Supplementary Information[33]). Next, the tip apex is positioned over a protrusion with the size of choice: the smaller, the stronger photon confinement is achieved.[17] Upon selecting a 1 nm high protrusion, we then record Raman spectra and the AFM cantilever phase kinetics with a gradual increase in the pumping intensity to 5.7 MW/cm$^2$ for 140 s. The optical heating is indicated by three observations: i) a temperature-dependent blueshift of the first-order VRS peak of Si at 521 cm$^{-1}$ (Fig. 2b), ii) high-frequency broadband inelastic ELS (Fig. 2d), exhibiting a nonlinear pump-dependent intensity behavior (Fig. 2c), and iii) a change in the AFM cantilever phase above the pumping intensity of 1.5 MW/cm$^2$ (Fig. 2e). The pump-dependent temperature of the tip apex is determined by a Raman-shift-based probe with an accuracy of 50 K (1800 grooves per mm grating) using Eq. S2 (Section II, the SI[33]). A calibration of the temperature probe was prior performed by temperature-dependent Raman measurements in the range from 25 to 600 °C using a heating stage (Linkam Scientific Model THMS600).

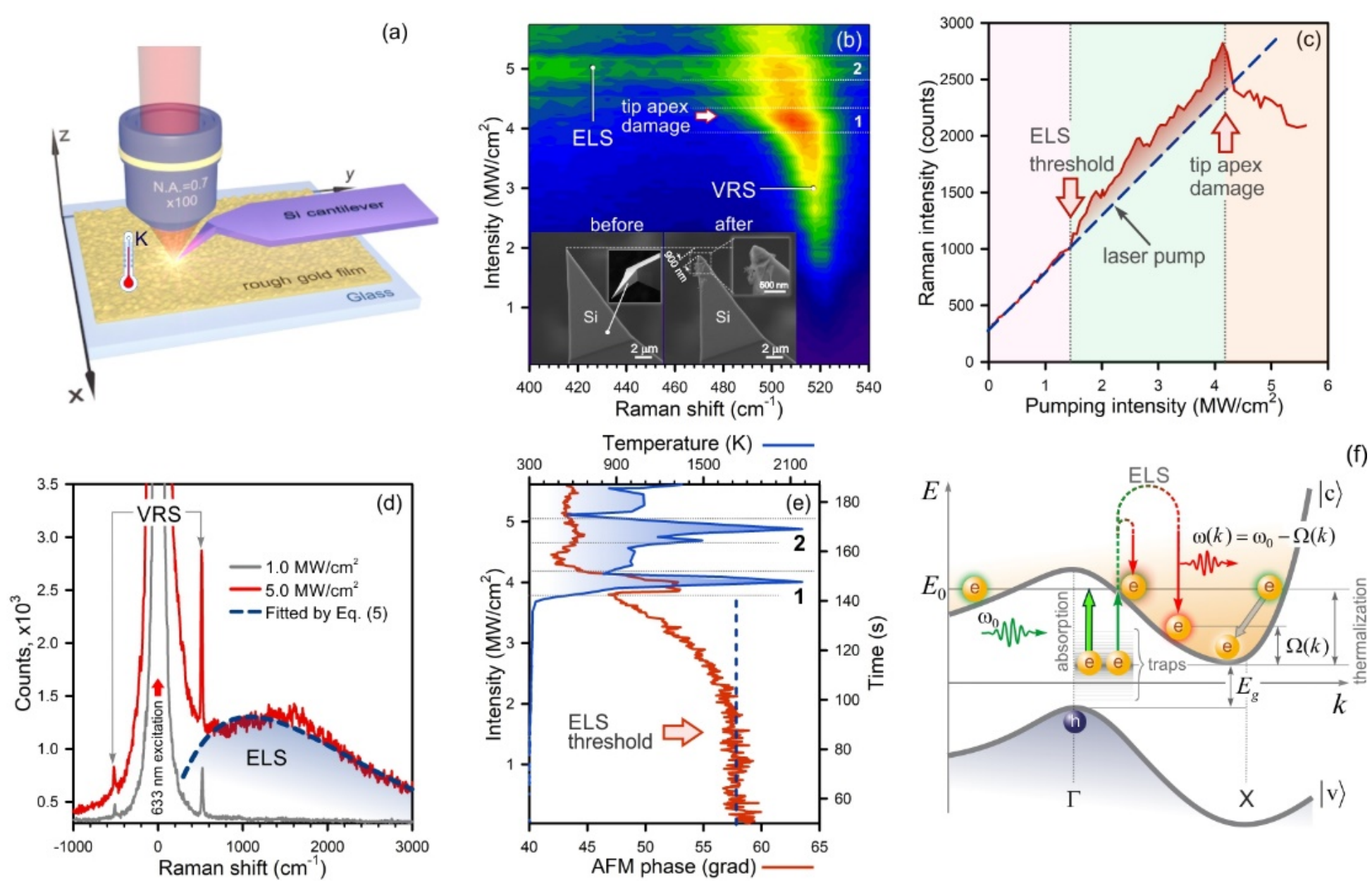

**Figure 2.** (a) Sketch of a TERS setup (upright configuration). (b) A Raman map of the AFM tip apex exposed to a focused laser light with different pumping intensity (the color indicates Raman intensity). The bottom inset displays SEM images of the AFM cantilever tips before and after laser impact with the intensity of 5 MW/cm$^2$. (c) A plot of the VRS intensity vs the pumping intensity. (d) Raman spectra taken at the intensity of 1 MW/cm$^2$ (gray) and 5 MW/cm$^2$ (red) in Fig. 2 (b). The dashed curve denotes a fit by Eq. (6). (e) A kinetics of the AFM cantilever phase (red) and a dependence of Raman-shift-calculated temperature vs the pumping intensity (blue). (f) The energy-momentum diagram for Si.

Upon turning the level of 4.2 MW/cm$^2$ (Fig. 2b and Fig. 2e), the temperature steeply climbs above 2000 K, making the tip apex melt. A close inspection of scanning electron microscopy (SEM) images of the tip apex (the inset in Fig. 2b), visualized before and after laser impact, confirms its destruction within 900 nm extent. However, the AFM cantilever phase starts to respond at 1.5 MW/cm$^2$ (red curve, Fig. 2e). The phase shift $\Delta\varphi \approx \frac{Q}{K}\frac{\partial F}{\partial z}$ (where $K$ is the spring constant, $Q$ is the quality factor) is mainly driven by the normal force gradient $\partial F/\partial z$ between the tip apex and the sample, that is greatly sensitive to temperature.[34,35] It is important to note that the temperature of the entire cantilever remains constant during our experiment, except the tip apex. It follows from the fact that the resonant frequency of the AFM cantilever, equal to 300 kHz, is fixed until the tip apex is damaged, the event marked with the arrow in Fig. 2b. In this figure, we highlight two regions labelled as '1' and '2' where giant temperature bursts are seen (Fig. 2e), accompanied by background emission (Fig. 2d). This observation is in good agreement with an increase in the Raman intensity at this level (Fig. 2c) that occurs due to ELS, as follows from Fig. 2d. Large temperature fluctuations above 4 MW/cm$^2$ (blue curve, Fig. 2e) are directly related to melting and a change in tip apex morphology. We conclude that the ELS mechanism holds promise for probing a local temperature at the hot spot.

Anomalous optical heating of spatially dispersive silicon is provided by near-field photons generated by a rough Au film (Fig. 2a). Clearly, the film itself should get heated as well. Due to the high thermal conductivity of the Au film, equal to 320 W/mK, the film is rapidly thermalized and, therefore, it is not heated.

In silicon, that is an indirect bandgap semiconductor, the optical absorption is driven by three-body photon-phonon-electron interaction. Alternatively, this process can be activated using a near-field photon carrying not only energy, but also momentum sufficient for indirect transitions.[17] Since the electronic density of states is maximal near the band edge,[30] absorption-based indirect transitions are limited because of energy band $k$-dispersion, shown in Fig. 2f. In general, one should consider optical transitions to those electronic states that correspond to energy $E_0$. Despite the fact that the transitions supported with large photon momenta are accessible, their contributions to absorption are little. In contrast, light scattering allows indirect transitions to all accessible states below energy $E_0$ near the conduction band edge. This leads to a broadband inelastic light scattering with frequency shifts: $\omega(k) = \omega_0 - \Omega(k)$ ($\omega_0$ is the pumping frequency, $\Omega(k)$ is a vibronic frequency associated with the energy band), directly observed in Fig. 2b and Fig. 2d. Eq. 5 can be used for fitting the high-frequency ELS peak at 5 MW/cm$^2$ (Fig. 2d), by using the regularized least squares method that yields the Penn energy $E_p = 136$ meV (1100 cm$^{-1}$). The net temperature rise of the tip apex is mainly determined by thermal conductivity of the tip shaft served as a heatsink, whereas heat exchange with the surroundings through its surface (Kapitza resistance[36]) is negligible. It is important to notice that the mechanism of inhomogeneous broadening of a central peak at $\omega(k) \approx \omega_0$ is related to low-energy ELS by a rough gold film that exists even at modest pump (Fig. 2d).[37,38]

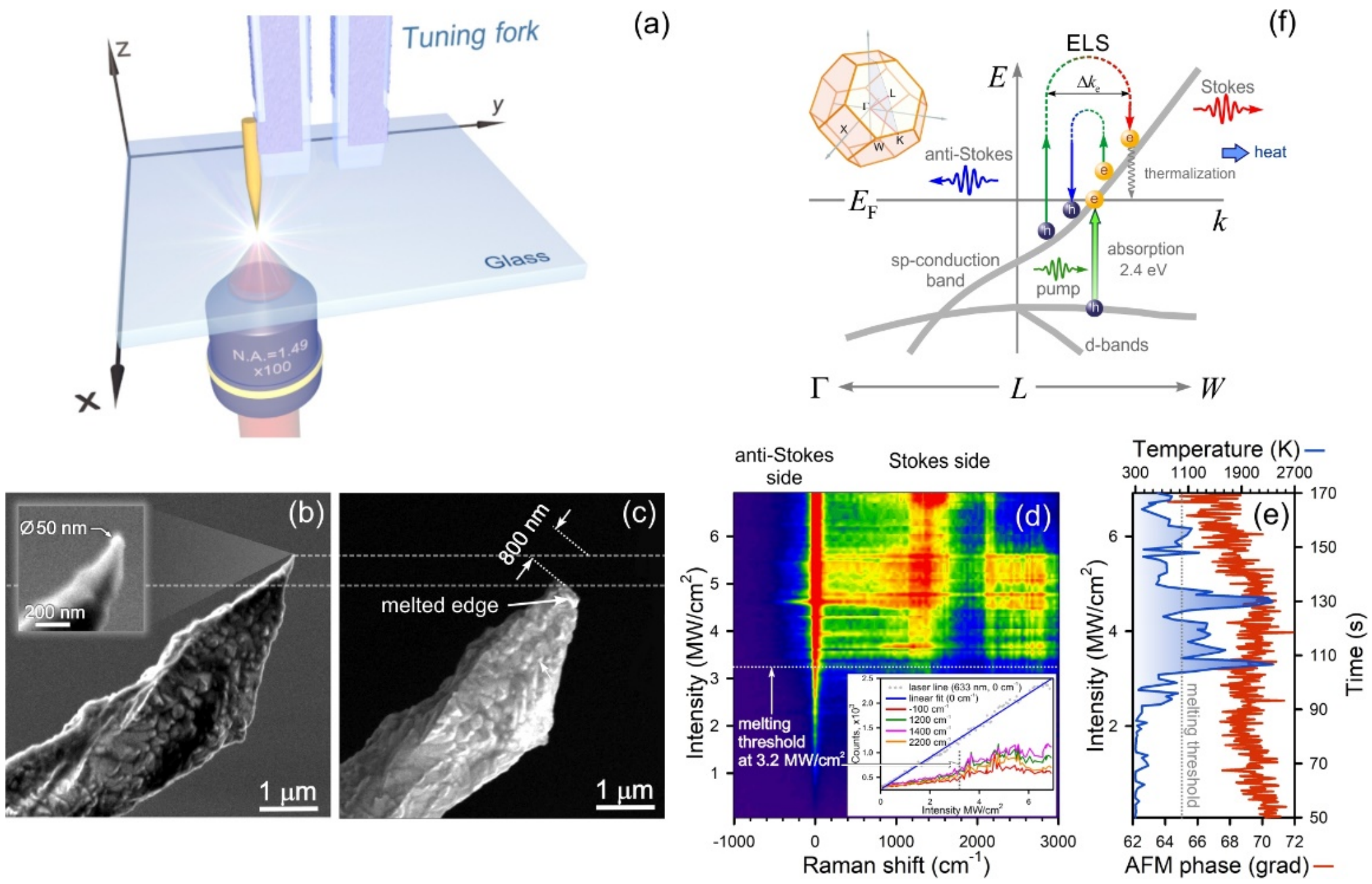

**Figure 3.** (a) Sketch of a TERS setup (inverted configuration). SEM images of a rough Au tips before (b) and after (c) laser illumination with the intensity of 5 MW/cm$^2$. (d) A Raman map of the Au tip upon optical heating. The inset shows a dependence of Raman intensity vs pumping intensity for different wavenumbers. (e) A kinetics of the Au tip phase (red) and a dependence of temperature (blue) vs pumping intensity when illuminated by a 633 nm laser light. (f) Schematic illustration of anti-Stokes/Stokes ELS with varying electron momentum $\Delta k_e$ and interband absorption in gold. The up-left inset shows the Brillouin zone for gold bulk.

Let us now consider the spatial confinement effect for metals. Fig. 3a shows a bottom-illumination TERS setup in which a rough gold tip glued to a quartz tuning fork oscillates in a plane parallel to the surface of bare coverslip with a frequency of 32 kHz. The inset in Fig. 3b displays the gold tip apex of 25 nm in curvature radius. Using the same protocol as for Si, we reveal optical melting of the Au tip apex within 800 nm extent. This paradoxical result is explained by the fact that the melting points of Si (1687 K) and Au (1337 K) differ by 30%, whereas their thermal conductivities (150 W/mK and 320 W/mK, respectively) are roughly two-fold distinct. In this case, a near-field photon is directly generated at the Au tip apex that is first heated and then destroyed. This effect indicates the fundamental role of expanded near-field photon momentum, enhancing light-matter interaction significantly.

We detect not only a low-frequency ELS signal often found in spatially confined metals, but also a broadband high-frequency ELS band (Fig. 3d). The optical melting occurs upon the intensity of 3.2 MW/cm$^2$. Like Si, the ELS intensity increases nonlinearly above this threshold, as seen in the inset of Fig. 3d. Since the Q-factor of the tuning fork is significantly worse than that of the AFM cantilever, the shear-force-microscopy phase is less sensitive to temperature (red curve in Fig. 3e). Following Sheldon et al, we used a combination of both Bose-Einstein and Fermi-Dirac statistics as a fitting function for estimating a temperature inside the Au tip apex by using the regularized least squares method:[39]

$$f_{\chi}(E,T) = \chi f_{FD}(E,T) + (1-\chi) f_{BE}(E,T), \tag{7}$$

where $\chi$ is a relative contribution of both to the distribution. The melting temperature is in good agreement with the intensity threshold value at which the ELS intensity starts to increase in Fig. 3d. Multiple temperature spikes are caused by dynamical destruction of the tip apex upon cw illumination impact (blue curve, Fig. 3e). Unlike a semiconductor, indirect optical transitions in spatially dispersive metals occur near the Fermi level, contributing to both anti-Stokes and Stokes signals. The temperature, calculated by using Eq. (7), was found from the anti-Stokes wing, displayed in Fig. S3 (Section III, the SI[33]). Interband absorption at 2.4 eV and above reduces the intensity of anti-Stokes scattering and, therefore, this method should be exploited with caution as temperature estimates can be biased. It is important to note that estimating temperature uncertainty is a challenging task because it requires *a priori* information on the energy band diagram, as implied by Eq. 6. In our case, the pump-dependent temperature of the gold tip apex is determined by an anti-Stokes probe with an accuracy of ca. 150 K (600 grooves per mm). Ultimately, the optical heating is determined by the thermalization of charge carries and depends on the slope of the sp-conduction band (Fig. 3f).

In this Letter, we have explored the optical heating of spatially dispersive semiconductors and metals driven by electronic light scattering rather than absorption. The underlying mechanism is indirect optical transitions caused by a change in the electron momentum. This phenomenon, often encountered in SERS/TERS experiments, is perceived as a parasitic background emission.[40–43] However, an optical nanoantenna (nanoparticle, quantum dot, defect, etc.) presents a system with strong spatial dispersion, generating broadband inelastic emission that depends on the excitation wavelength and extends over several thousands of $cm^{-1}$.[5] This emission, that is specifically the ELS, carries important information on the spatial structure of nonlocal media, and does not depend on their chemical composition. The ELS mechanism alters an equilibrium of the electron system, leading to an increase in the real refractive index and dc-conductivity.[3] More importantly, it can heat up nonlocal media under cw illumination owing to their energy band $k$-dispersion. This phenomenon is based on expanded near-field photon momentum that provides enhanced interaction between light and matter at the nanoscale. This means that media with strong spatial dispersion can be heated mainly due to the delocalization of optical near-fields.[6]

Nonlocal photonics is critical for engineering white light-emitting diodes and mirrorless lasers at room-temperature,[3] silicon solar cells with efficiency beyond the Shockley-Queisser limit[44,45] and optically transparent conducting electrodes.[29] A special attention should be paid to biological systems that are highly heterogeneous. ELS opens the route to optical transparency of biological tissues[1] and optical recognition of peptide and protein conformations.[46] A promising feature of spatially dispersive media is their optical heating under sub-band pumping. It brings optical heating into sharper focus for thermo-selective treatment of neurodegenerative and oncological diseases.[27]

**Disclosures.** The authors declare no conflicts of interest.

**ACKNOWLEDGEMENT**

The authors gratefully thank A.M. Rogov for carrying out SEM measurements and A.B. Shubin for fruitful discussions.

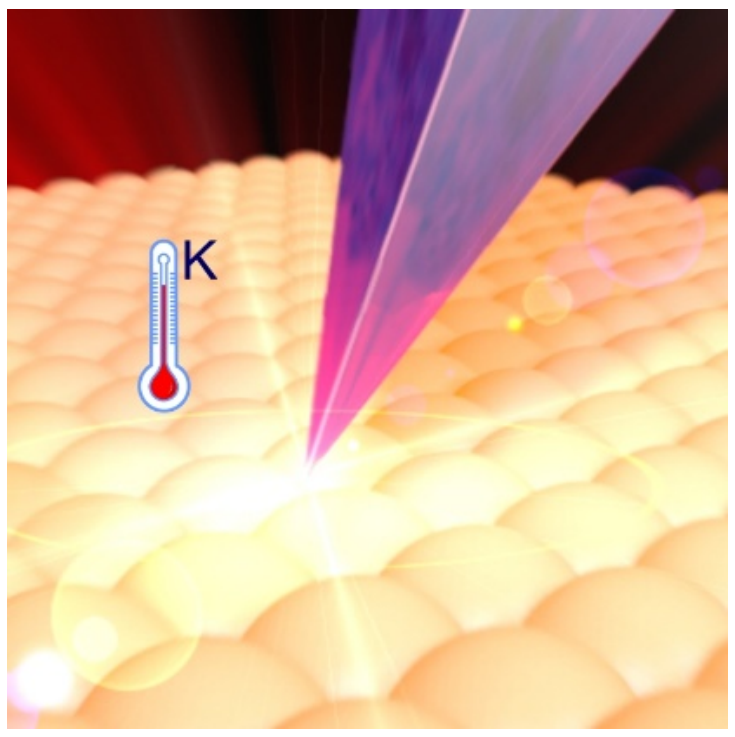

TOC image